\documentstyle[12pt]{article}

\font\twlgot =eufm10 scaled \magstep1 \font\egtgot =eufm8
\font\sevgot =eufm7 \font\twlmsb =msbm10 scaled \magstep1
\font\egtmsb =msbm8 \font\sevmsb =msbm7

\newfam\gotfam
\def\pgot{\fam\gotfam\twlgot}
\textfont\gotfam\twlgot \scriptfont\gotfam\egtgot
\scriptscriptfont\gotfam\sevgot
\def\got{\protect\pgot}
\newfam\msbfam
\textfont\msbfam\twlmsb \scriptfont\msbfam\egtmsb
\scriptscriptfont\msbfam\sevmsb
\def\Bbb{\protect\pBbb}
\def\pBbb{\relax\ifmmode\expandafter\Bb\else\typeout{You cann't use
Bbb in text mode}\fi}
\def\Bb #1{{\fam\msbfam\relax#1}}

\newcommand{\gd}{{\got d}}

\newcommand{\ccG}{{\got g}}

\def\thebibliography#1{\section*{References}\list
  {[\arabic{enumi}]}{\settowidth\labelwidth{#1}\leftmargin\labelwidth
    \advance\leftmargin\labelsep
    \usecounter{enumi}}
    \def\newblock{\hskip .11em plus .33em minus .07em}
    \sloppy\clubpenalty4000\widowpenalty4000
    \sfcode`\.=1000\relax}

\def\op#1{\mathop{\fam0 #1}\limits}

\newcommand{\id}{{\rm Id\,}}

\newcommand{\nm}[1]{|{#1}|}
\newcommand{\beq}{\begin{equation}}
\newcommand{\eeq}{\end{equation}}
\newcommand{\ben}{\begin{eqnarray}}
\newcommand{\een}{\end{eqnarray}}
\newcommand{\be}{\begin{eqnarray*}}
\newcommand{\ee}{\end{eqnarray*}}
\newcommand{\bea}{\begin{eqalph}}
\newcommand{\eea}{\end{eqalph}}

\newcommand{\cR}{{\cal R}}

\newcommand{\cH}{{\cal H}}

\newcommand{\cS}{{\cal S}}

\newcommand{\sG}{{\cal G}}

\newcommand{\al}{\alpha}

\newcommand{\bt}{\beta}
\newcommand{\dl}{\delta}
\newcommand{\la}{\lambda}
\newcommand{\La}{\Lambda}
\newcommand{\f}{\phi}
\newcommand{\om}{\omega}

\newcommand{\g}{\gamma}

\newcommand{\vf}{\varphi}

\newcommand{\di}{{\rm dim\,}}
\newcommand{\pr}{{\rm pr}}

\newcommand{\si}{\sigma}
\newcommand{\Si}{\Sigma}
\newcommand{\w}{\wedge}

\newcommand{\wh}{\widehat}
\newcommand{\ol}{\overline}
\newcommand{\dr}{\partial}
\newcommand{\ar}{\op\longrightarrow}

\newcommand{\ot}{\otimes}

\newcommand{\ve}{\varepsilon}

\newcounter{theorem}
\newcounter{remark}
\newcounter{proposition}
\newcounter{lemma}
\newcounter{corollary}
\newcounter{definition}

\def\theremark{\arabic{remark}}

\def\thedefinition{\arabic{definition}}

\newenvironment{theo}{\refstepcounter{definition} \medskip
\noindent{\bf Theorem \thedefinition.}}{\medskip}

\newenvironment{defi}{\refstepcounter{definition} \medskip
\noindent{\bf Definition \thedefinition.}}{\medskip}

\newcommand{\mar}[1]{}

\begin{document}
\hbox{}

\begin{center}

{\large\bf SUPERMETRICS ON SUPERMANIFOLDS}
\bigskip

{\sc G.SARDANASHVILY}

{\it Department of Theoretical Physics, Moscow State University,
117234, Moscow, Russia}
%\\ sardanashvi@phys.msu.ru

\end{center}

\begin{small}

\bigskip
By virtue of the well-known theorem, a structure Lie group $K$ of
a principal bundle $P\to X$ is reducible to its closed subgroup
$H$ iff there exists a global section of the quotient bundle
$P/K\to X$. In gauge theory, such sections are treated as Higgs
fields, exemplified by pseudo-Riemannian metrics on a base
manifold $X$. Under some conditions, this theorem is extended to
principal superbundles in the category of $G$-supermanifolds.
Given a $G$-supermanifold $M$ and a graded frame superbundle over
$M$ with a structure general linear supergroup, a reduction of
this structure supergroup to an orthgonal-symplectic supersubgroup
is associated to a supermetric on a $G$-supermanifold $M$.
\end{small}

\section{Introduction}

Let $\pi:P\to X$ be a principal smooth bundle with a structure Lie
group $K$. One says that its structure group $K$ is reducible to a
closed (consequently, Lie) subgroup $H$ of $K$, if there exists an
$H$-principal subbundle of $P$. The following well-known theorem
holds \cite{ste}.

\begin{theo} \label{g00} \mar{g00} There is one-to-one
correspondence between the reduced $H$-principal subbundles $P^h$
of $P$ and the global sections $h$ of the quotient bundle $P/H\to
X$.
\end{theo}

In gauge theory on a principal bundle $P\to X$, these sections $h$
are regarded as classical Higgs fields \cite{sard92,higgs} In
particular, let $P=LX$ be the $GL(n,\Bbb R)$-principal bundle of
linear frames in the tangent bundle $TX$ of $X$ (n=\di X). If
$H=O(k,n-k)$, then a global section of the quotient bundle
$LX/O(k,n-k)$ is a pseudo-Riemannian metric on $X$. For instance,
this is the case of a gravitational field in gauge gravitation
theory \cite{iva,tmf}.

SUSY gauge theory including supergravity is mainly developed as a
Yang--Mills type theory with spontaneous breaking of
supersymmetries \cite{west,wess,binet}. However, supergravity
introduced in SUSY gauge theory has no geometric feature as a
supermetric \cite{iva86,del}. Therefore, our goal here is to
extend Theorem \ref{g00} to principal superbundles.

It should be emphasized that there are different notions of a
supermanifold, Lie supergroup and superbundle. Let us mention a
definition of a super Lie group as a Harish--Chandra pair of a Lie
group and a super Lie algebra \cite{del,carm}. We are not
concerned with graded manifolds, graded Lie algebras and principal
bundles \cite{alm,boy,stavr}. A problem is that the tangent bundle
of a graded manifold fails to be a graded manifold, but it is
locally a Ne'eman-Quillen superbundle \cite{book00} (see
\cite{zirn,aleks,goert} for different definitions of a supermetric
on graded manifolds). It should be also emphasized that graded
manifolds are not supermanifolds in a strict sense. Both graded
manifolds and supermanifolds are described in terms of sheaves of
graded commutative algebras \cite{bart}. However, graded manifolds
are characterized by sheaves on smooth manifolds, while
supermanifolds are constructed by gluing of sheaves on supervector
spaces.

We here restrict our consideration to supermanifolds over
Grassmann algebras $\La$ of finite rank. This is the case of
smooth ($GH^\infty$-, $H^\infty$-, $G^\infty$-) supermanifolds and
$G$-supermanifolds \cite{bart}. By analogy with manifolds, smooth
supermanifolds are constructed by gluing of open subsets of
supervector spaces $B^{n,m}$ endowed with the Euclidean topology.
If a supervector space $B^{n,m}$ is provided with the
non-Hausdorff DeWitt topology and a smooth or $G$-supermanifold
admits an appropriate atlas, it is a DeWitt supermanifold.

In a general setting, one considers supermanifolds over the so
called Arens--Michael algebras of Grassmann origin. They are $R$-
and $R^\infty$-supermanifolds which obey a certain set of axioms
in order to be most suitable for supergeometry
\cite{roth,bart93,bruz99}. In the case of a finite Grassmann
algebra, the category of $R^\infty$-supermanifolds is equivalent
to the category of $G$-supermanifolds. In comparison with smooth
supermanifolds, $G$-supermanifolds have some important advantages
from the differential geometric viewpoint \cite{bart}. Firstly,
derivations of the structure sheaf of a $G$-supermanifold $\wh M$
constitute a locally free sheaf, which is the structure sheaf of
some $G$-superbundle $T\wh M$ regarded as a tangent superbundle of
$\wh M$. Secondly, the category of $G$-supervector bundles is
equivalent to a certain category of locally free sheaves of finite
rank just as it takes place in the case of smooth vector bundles.

Therefore, we consider Lie supergroups and principal superbundles
in the category of $G$-supermanifolds. Our goal here is the
following extension of Theorem \ref{g00} to principal
$G$-superbundles.

\begin{theo} \label{g20} \mar{g20}
Let $\wh P\to \wh M$ be a principal $G$-superbundle with a
structure $G$-Lie supergroup $\wh K$, and let $\wh H$ be a closed
$G$-Lie supersubgroup of $\wh K$ such that $\wh K\to \wh K/\wh H$
is a principal superbundle. There is one-to-one correspondence
between the principal $G$-supersubbundles of $\wh P$ with the
structure $G$-Lie supergroup $\wh H$ and the global sections of
the quotient superbundle $\wh P/\wh H\to \wh M$ with the typical
fiber $\wh K/\wh H$.
\end{theo}

A key point is that underlying spaces of $G$-supermanifolds are
smooth real manifolds, but possessing very particular transition
functions. Therefore, the condition of local triviality of the
quotient $\wh K\to \wh K/\wh H$ is rather restrictive. It is
satisfied in the case of DeWitt $G$-supermanifolds \cite{klepp}
and in the most interesting case for applications when $\wh K$ is
a supermatrix group and $\wh H$ is its Cartan supersubgroup. For
instance, let $\wh P=L\wh M$ be a principal superbundle of graded
frames in the tangent superspaces over a supermanifold $\wh M$ of
even-odd dimensione $(n,2m)$. If its structure general linear
supergroup $\wh K=\wh{GL}(n|2m; \La)$ is reduced to the
orthgonal-symplectic supersubgroup $\wh H=\wh{OS}p(n|m;\La)$, one
can think of the corresponding global section of the quotient
bundle $L\wh M/\wh H\to \wh M$ as being a supermetric on $\wh M$.

\section{Supermanifolds}

Let $V$ be a real vector space. Its exterior algebra
\be \La=\La_0\oplus\La_1=\w V=\Bbb R\op\oplus_{k=1} \op\w^k V
=(\Bbb R\op\oplus_{k=1} \op\w^{2k} V)\oplus(\op\oplus_{k=1}
\op\w^{2k-1} V)
\ee
is a graded commutative ring of rank $N=\di V$. We consider
Grassmann algebras of this type. A Grassmann algebra admits the
splitting
\mar{+11}\beq
\La=\Bbb R\oplus\cR =\Bbb R\oplus\cR_0\oplus\cR_1=\Bbb R\oplus
(\La_1)^2\oplus \La_1 \label{+11}
\eeq
where $\cR$ is the ideal of nilpotents of $\La$. It is a unique
maximal ideal of $\La$, i.e., $\La$ is a local ring. The
corresponding projections $\si:\La\to\Bbb R$ and $s:\La\to\cR$ are
called the body and soul maps, respectively. Given a basis
$\{c^i\}$ for the vector space $V$, elements of the Grassmann
algebra take the form
\be
a=\op\sum_{k=0}^N \frac{1}{k!}a_{i_1\cdots i_k}c^{i_1}\cdots
c^{i_k}.
\ee
A Grassmann algebra $\La$ is a graded commutative Banach ring with
respect to the norm
\be
\|a\|_\La=\op\sum_{k=0}^N \op\sum_{(i_1\cdots i_k)}\frac{1}{k!}
\nm{a_{i_1\cdots i_k}}.
\ee
This norm provides $\La$ with the Euclidean topology of a
$2^N$-dimensional real vector space.

Let $B=B_0\oplus B_1=\Bbb R^n\oplus\Bbb R^m$ be an
$(n,m)$-dimensional graded vector space. Given a Grassmann algebra
$\La$ of rank $N$, the $\La$-envelope of $B$ is a free graded
$\La$-module
\mar{+70}\beq
B^{n|m}=\La B=(\La B)_0\oplus (\La B)_1=(\La_0^n \oplus
\La^m_1)\oplus (\La_1^n\oplus \La_0^m), \label{+70}
\eeq
of rank $n+m$. It is called the superspace. Its even part
\mar{g11}\beq
 B^{n,m}= \La_0^n \oplus \La^m_1 \label{g11}
\eeq
is a $\La_0$-module called the $(n,m)$-dimensional supervector
space. In accordance with the decomposition (\ref{+11}), any
element $q\in B^{n,m}$ is uniquely split as
\be
 q=x+y=(\si(x^i) + s(x^i))e^0_i + y^je^1_j,
\ee
where $\{e^0_i,e^1_j\}$ is a basis for $B$ and $\si(x^i)\in \Bbb
R$, $s(x^i)\in \cR_0$, $y^j\in\cR_1$. The corresponding body and
soul maps read
\be
\si^{n,m}: B^{n,m}\to \Bbb R^n, \qquad s^{n,m}: B^{n,m}\to
\cR^{n,m}=\cR^n_0\oplus \cR^m_1.
\ee
A supervector space $B^{n,m}$ is provided with the Euclidean
topology of a $2^{N-1}(n+m)$-dimensional real vector space.

The superspace $B^{n|m}$ seen as a $\La_0$-module is isomorphic to
the supervector space $B^{n+m,n+m}$. Any $\La$-module endomorphism
of $B^{n| m}$ is represented by an $(n+ m)\times (n+m)$-matrix
\mar{+200}\beq
L=\left(
\begin{array}{cc}
L_1 & L_2 \\
L_3 & L_4
\end{array}
\right) \label{+200}
\eeq
with entries taking values in $\La$. It is called a supermatrix. A
supermatrix $L$ is even (resp. odd) if $L_1$ and $L_4$ have even
(resp. odd) entries, while $L_2$ and $L_3$ have the odd (resp.
even) ones. Endowed with this gradation, supermatrices
(\ref{+200}) make up a $\La$-graded algebra. A supermatrix $L$
(\ref{+200}) is invertible iff the real matrix $\si(L)$ is
invertible. Invertible supermatrices constitute a general linear
graded group $GL(n|m;\La)$.

Turn now to the notion of a superfunction. Let $B^{n,m}$ be a
supervector space (\ref{g11}), where $\La$ is a Grassmann algebra
of rank $0<m\leq N$. Let $\La'$ be a Grassmann subalgebra of $\La$
of rank $N'$ whose basis $\{c^a\}$ is a subset of a basis for
$\La$. Furthermore, $\La$ is regarded as a $\La'$-algebra. Given
an open subset $U\subset \Bbb R^n$, let us consider a
$\La'$-valued graded function
\mar{+12}\beq
f(z)=\op\sum_{k=0}^{N'} \frac1{k!}f_{a_1\ldots
a_k}(z)c^{a_1}\cdots c^{a_k}, \label{+12}
\eeq
on $U$ with smooth real coefficients $f_{a_1\cdots a_k}\in
C^\infty(U)$. It is prolonged onto $(\si^{n,0})^{-1}(U)\subset
B^{n,0}$ as the Taylor series
\mar{+14}\beq
f(x)=  \op\sum_{k=0}^{N'} \frac1{k!}\left[
 \op\sum_{p=0}^N\frac{1}{p!}\frac{\dr^qf_{a_1\ldots a_k}}{\dr
z^{i_1}\cdots \dr z^{i_p}}(\si(x))s(x^{i_1})\cdots
s(x^{i_p})\right]c^{a_1}\cdots c^{a_k}. \label{+14}
\eeq
Then a superfunction $F(q)=F(x,y)$ on $(\si^{n,m})^{-1}(U)\subset
B^{n,m}$ is defined as the sum
\mar{+13}\beq
F(x,y)= \op\sum_{r=0}^N \frac1{r!} f_{j_1\ldots
j_r}(x)y^{j_1}\cdots y^{j_r}, \label{+13}
\eeq
where $f_{j_1\ldots j_r}(x)$ are graded functions (\ref{+14}). The
representation of a superfunction $F(x,y)$ by the sum (\ref{+13})
however need not be unique. The germs of superfunctions
(\ref{+13}) constitute the sheaf $S_{N'}$ of graded commutative
$\La'$-algebras on $B^{n,m}$. At the same time, one can think of a
superfunction (\ref{+13}) as being a smooth map of a
$2^{N-1}(n+m)$-dimensional Euclidean space $B^{n,m}$ to the
$2^N$-dimensional one $\La$.

Using the representation (\ref{+13}), one can define derivatives
of superfunctions as follows. Given a superfunction $f(x)$
(\ref{+14}) on $B^{n,0}$, its derivative with respect to an even
argument $x^i$ is defined in a natural way as
\mar{+16}\beq
\dr_if(x)= \op\sum_{k=0}^{N'} \frac1{k!}\left[
 \op\sum_{p=0}^N\frac{1}{p!}\frac{\dr^{p+1}f_{a_1\ldots a_k}}{\dr
z^i\dr z^{i_1}\cdots \dr z^{i_p}}(\si(x))s(x^{i_1})\cdots
s(x^{i_p})\right]c^{a_1}\cdots c^{a_k}. \label{+16}
\eeq
This even derivative is extended to superfunctions $F$ on
$B^{n,m}$ in spite of the fact that the representation (\ref{+13})
is not unique. However, the definition of odd derivatives of
superfunctions is more intricate.

Let $S_{N'}^0\subset S_{N'}$ be the subsheaf of superfunctions
$F(x,y)=f(x)$ (\ref{+14}) independent of the odd arguments $y^j$.
Let $\w\Bbb R^m$ be the Grassmann algebra generated by the
elements $(a^1,\ldots, a^m)$. The expression (\ref{+13}) implies
the sheaf epimorphism
\mar{+15}\be
\la: S^0_{N'}\ot\w\Bbb R^m \to S_{N'}, \qquad \op\sum_{r=0}^m
\frac1{r!} f_{j_1\ldots j_r}(x)\ot(a^{j_1}\cdots a^{j_r})\to
\op\sum_{r=0}^m \frac1{r!} f_{j_1\ldots j_r}(x)y^{j_1}\cdots
y^{j_r}. \label{+15}
\ee
This epimorphism is injective and, consequently, is an isomorphism
iff
\mar{+20}\beq
N-N'\geq m. \label{+20}
\eeq
If the condition (\ref{+20}) holds, the representation of any
superfunction $F$ by the sum (\ref{+13}) is unique, and $F$ is an
image of some section $f_\iota\ot a^\iota$ of the sheaf
$S^0_{N'}\ot\w\Bbb R^m$. Then the odd derivative of $F$ is defined
as
\be
\frac{\dr}{\dr y^j}(\la(f_\iota\ot a^\iota))=\la (f_\iota\ot
\frac{\dr}{\dr a^j}(a^\iota)).
\ee
This definition is consistent only if $\la$ is an isomorphism. If
otherwise, there exists a non-vanishing element $f_\iota\ot
a^\iota$ such that $\la(f_\iota\ot a^\iota)=0$, but $\la
(f_\iota\ot \dr_j(a^\iota))\neq 0$.

With the condition (\ref{+20}), one classifies superfunctions as
follows \cite{bart,rog}.

(i) If the condition (\ref{+20}) is satisfied, superfunctions
(\ref{+13}) are called $GH^\infty$-superfunctions.

(ii) If $N'=0$, the condition (\ref{+20}) holds, and we deal with
$H^\infty$-superfunctions
\mar{+41}\beq
F(x,y)=\op\sum_{r=0}^m \frac1{r!}\left[
 \op\sum_{p=0}^N\frac{1}{p!}\frac{\dr^qf_{j_1\ldots j_r}}{\dr
z^{i_1}\cdots \dr z^{i_p}}(\si(x))s(x^{i_1})\cdots
s(x^{i_p})\right]y^{j_1}\cdots y^{j_r}, \label{+41}
\eeq
where $f_{j_1\ldots j_r}$ are smooth real functions.

(iii) If $N'=N$, the inequality (\ref{+20}) is not satisfied,
unless $m=0$. This is the case of $G^\infty$-superfunctions.

Superfunctions of these three types are called smooth
superfunctions. They however are effected by serious
inconsistencies. Firstly, the odd derivatives of
$G^\infty$-superfunctions are ill defined. Secondly, a space of
values of $GH^\infty$-superfunctions changes from point to point
because a Grassmann algebra $\La$ fails to be a free
$\La'$-module. Though the $H^\infty$-superfunctions are free of
these defects, they are rather particular, namely, they are even
on $B^{n,0}$ and real on $\Bbb R^n\subset B^{n,0}$. The notion of
$G$-superfunctions overcomes these difficulties.

Let $\sG\cH_{N'}$ denote the sheaf of $GH^\infty$-superfunctions
on a supervector space $B^{n,m}$. Let us consider the sheaf of
graded commutative $\La$-algebras
\be
\sG_{N'}=\sG\cH_{N'}\op\ot_{\La'} \La.
\ee
There is its evaluation morphism
\be
\dl:\sG_{N'}\ni F\ot a\mapsto Fa\in C^{0\La}(B^{n,m})
\ee
to the ring $C^{0\La}(B^{n,m})$ of continuous $\La$-valued
functions on $B^{n,m}$. It is an epimorphism onto the sheaf
$\sG^\infty$ of $G^\infty$-superfunctions on $B^{n,m}$. A key
point is that, for any two integers $N'$ and $N''$ satisfying the
condition (\ref{+20}), there is the canonical isomorphism of
sheaves of graded commutative $\La$-algebras $\sG_{N'}$ and
$\sG_{N''}$. Therefore, given the sheaf $\cH^\infty$ of
$H^\infty$-superfunctions $F$ (\ref{+41}) on a supervector space
$B^{n,m}$, it suffices to consider the canonical sheaf
\mar{g23}\beq
\sG_{n,m}=\cH^\infty\ot \La, \qquad \dl: \sG_{n,m}\to
C^{0\La}_{B^{n,m}}, \label{g23}
\eeq
of graded commutative $\La$-algebras on $B^{n,m}$. Its sections
are called $G$-superfunctions.

It is important from the geometric viewpoint that the sheaf
$\gd\sG_{n,m}$ of graded derivations of the sheaf $\sG_{n,m}$
(\ref{g23}) is a locally free sheaf of $\sG_{n,m}$-modules of rank
$(n,m)$. On any open set $U\subset B^{n,m}$, the
$\sG_{n,m}(U)$-module $\gd\sG_{n,m}(U)$ is generated by the
derivations $\dr/\dr x^i$, $\dr/\dr y^j$ which act on
$\sG_{n,m}(U)$ by the rule.
\mar{+83}\beq
\frac{\dr}{\dr x^i}(F\ot a)=\frac{\dr F}{\dr x^i}\ot a, \qquad
\frac{\dr}{\dr y^j}(F\ot a)=\frac{\dr F}{\dr y^j}\ot a.
\label{+83}
\eeq

\begin{defi} \mar{g31} \label{g31}
Given two open subsets $U$ and $V$ of a supervector space
$B^{n,m}$, a map $\f: U\to V$ is called supersmooth if it is a set
of $n+m$ smooth superfunctions. A Hausdorff paracompact
topological space $M$ is said to be an $(n,m)$-dimensional smooth
supermanifold if it admits an atlas
\be
\Psi=\{U_\zeta,\f_\zeta\}, \qquad \f_\zeta: U_\zeta\to B^{n,m}
\ee
such that the transition functions $\f_\zeta\circ\f_\xi^{-1}$ are
supersmooth. If transition functions are $GH^\infty$- $H^\infty$-,
or $G^\infty$-superfunctions, one deals with $GH^\infty$-
$H^\infty$-, or $G^\infty$-supermanifolds, respectively.
\end{defi}

By virtue of Definition \ref{g31}, any smooth supermanifold of
dimension $(n,m)$ also carries a structure of a smooth real
manifold of dimension $2^{N-1}(n+m)$, whose atlas however
possesses rather particular transition functions. Therefore, it
may happen that non-isomorphic smooth supermanifolds are
diffeomorphic as smooth manifolds.

Similarly to the case of smooth manifolds, Definition \ref{g31} of
smooth supermanifolds is equivalent to the following one
\cite{book00,bart}.

\begin{defi} \mar{g32} \label{g32}
A smooth supermanifold is a graded local-ringed space $(M,S)$ with
an underlying topological space $M$ and the structure sheaf $S$
which is locally isomorphic to the graded local-ringed space
$(B^{n,m},\cS)$, where $\cS$ is one of the sheaves of smooth
superfunctions on $B^{n,m}$.
\end{defi}

By a morphism of smooth supermanifolds is meant their morphism as
local-ringed spaces
\be
(\vf,\Phi): (M,S)\to (M',S'), \quad \vf: M\to M', \quad
\Phi(S')=(\vf_*\circ\vf^*)(S')\subset \vf_*(S),
\ee
where $\vf_*(S)$ is the direct image of a sheaf and $\vf^*S'$ is
the inverse (pull-back) one. The condition $\vf^*(S')\subset S$
implies that $\vf:M\to M'$ is a smooth map.

The notion of a $G$-supermanifold follows Definition \ref{g32} of
smooth supermanifolds.

\begin{defi} \mar{g33} \label{g33}
A $G$-supermanifold is a graded local-ringed space $\wh
M=(M,\sG_M)$ satisfying the following conditions:

(i) $M$ is a Hausdorff paracompact topological space;

(ii) $(M,\sG_M)$ is locally isomorphic to the graded local-ringed
space $(B^{n,m},\sG_{n,m})$, where $\sG_{n,m}$ is the sheaf of
$G$-superfunctions on $B^{n,m}$;

(iii) there exists an evaluation morphism $\dl:\sG_M\to
C^{0\La}_M$ to the sheaf $C^{0\La}_M$ of continuous $\La$-valued
functions on $M$ which is locally compatible to the evaluation
morphism (\ref{g23}).
\end{defi}

In particular, the triple $\wh B^{n,m}=(B^{n,m},\sG_{n,m},\dl)$,
where $\dl$ is the evaluation morphism (\ref{g23}), is called the
standard $G$-supermanifold.

Any $GH^\infty$-supermanifold $(M,GH^\infty_M)$ with the structure
sheaf $\sG\cH^\infty_M$ is naturally extended to the
$G$-supermanifold $(M,\sG_M=\sG\cH^\infty_M\ot \La)$. Every
$G$-supermanifold defines the underlying $G^\infty$-supermanifold
$(M,\sG^\infty_M=\dl(\sG_M))$. Therefore, the underlying space $M$
of a $G$-supermanifold $(M,\sG_M)$ is provided with the structure
of a real smooth manifold of dimension $2^{N-1}(n+m)$. However, it
may happen that non-isomorphic $G$-supermanifolds have
diffeomorphic underlying smooth manifolds.

Given a $G$-supermanifold $(M,\sG_M)$, the ring $\sG_M(M)$ of
$G$-superfunctions on $M$ becomes a Fr\'echet algebra with respect
to the topology of uniform convergence of derivatives of any order
defined by the family of seminorms
\be
p_{l,K}(F)=\op\max_{q\in K, (\al_1+\cdots +\al_k)\leq l}
\|\dl((\frac{\dr}{\dr x^{i_1}})^{\al_1}\cdots (\frac{\dr}{\dr
x^{i_k}})^{\al_k}\op\sum_{r=0}^m\op\sum_{(j_1\ldots
j_r)}\frac{\dr}{\dr y^{j_1}}\cdots \frac{\dr}{\dr
y^{j_r}}F)(q)\|_\La,
\ee
where $l\in \Bbb N$ and $K$ runs over compact subsets of $M$.
Given  the standard $G$-supermanifold $(B^{n,m},\sG_{n,m})$ and an
open $U\subset B^{n,m}$, there is an isomertical isomorphism of
Fr\'echet algebras
\be
\sG_{n,m}(U)\equiv C^{\infty\La}(\si^{n,m}(U))\ot\w\Bbb R^m=
C^\infty(\si^{n,m}(U))\ot\La\ot\w\Bbb R^m,
\ee
where $C^\infty(\Bbb R^m)\ot\La\ot\w\Bbb R^m$ is provided with the
corresponding topology of uniform convergence of derivatives of
any order. As a consequence, the evaluation morphism $\dl$
(\ref{g23}) takes its values into the sheaf
$C^{\infty\La}_{B^{n,m}}$ of smooth $\La$-valued functions on
$B^{n,m}$. Accordingly, the evaluation morphism of a
$G$-supermanifold $(M,\sG_M)$ is $\dl:\sG_M\to C^{\infty\La}_M$.

By a morphism of $G$-supermanifolds is meant their morphism as
local-ringed spaces
\mar{g35}\beq
\wh\vf= (\vf,\Phi): (M,\sG_M)\to (M',\sG_{M'}) \label{g35}
\eeq
over the pull-back morphism $(\vf, \vf_*\circ\vf^*)$ of the
underlying $G^\infty$-supermanifolds. It follows that $\vf:M\to
M'$ is a smooth map of underlying smooth manifolds. Note that a
map $\vf$ is not sufficient on its own in order to determine an
even sheaf morphism $\Phi$, and additional specification $\Phi$ is
needed.

A $G$-morphism (\ref{g35}) is said to be a monomorphism (resp.
epimorphism) if $\vf$ is injective (resp. surjective) and $\Phi$
is an epimorphism (resp. monomorphism). In particular, a
$G$-monomorphism $\wh\vf:\wh M'\to\wh M$ is called a
supersubmanifold if a morphism of underlying manifolds $\vf:M'\to
M$ is a submanifold, i.e., an injective immersion. If $\vf$ is
imbedding, a supersubmanifold is called imbedded.

\section{Superbundles}

As was mentioned above, we consider superbundles in the category
of $G$-supermanifolds \cite{book00,bart}. Therefore, we start with
the definition of a product of two $G$-supermanifolds.

Let $(B^{n,m},\sG_{n,m})$ and $(B^{r,s},\sG_{r,s})$ be two
standard $G$-supermanifolds. Given open sets $U\subset B^{n,m}$
and $V\subset B^{r,s}$, we consider the presheaf
\mar{+67}\beq
U\times V\to \sG_{n,m}(U)\wh\ot \sG_{r,s}(V) \label{+67}
\eeq
where $\wh\ot$ denotes the tensor product of modules completed in
the Grothendieck topology. This presheaf yields the structure
sheaf $\sG_{n+r,m+s}$ of the standard $G$-supermanifold $\wh
B^{n+r,m+s}$. This construction is generalized to arbitrary
$G$-supermanifolds as follows.

Let  $(M,\sG_M)$ and $(M',\sG_{M'})$ be two $G$-supermanifolds of
dimensions $(n,m)$ and $(r,s)$, respectively. Their product
$(M,\sG_M) \times(M',\sG_{M'})$ is defined as the graded
local-ringed space $(M\times M', \sG_M\wh\ot \sG_{M'})$, where
$\sG_M\wh\ot \sG_{M'}$ is the sheaf constructed from the presheaf
\be
U\times U'\to \sG_M(U)\wh\ot \sG_{M'}(U')
\ee
for any open subsets $U\subset M$ and $U'\subset M'$. This product
is a $G$-supermanifold of dimension $(n+r,m+s)$ provided with the
evaluation morphism
\be
\dl: \sG_M\wh\ot \sG_{M'}\to C^{\infty\La}_{M\times M'}.
\ee
There are the canonical $G$-epimorphisms
\be
\wh\pr_1:(M,\sG_M) \times(M',\sG_{M'})\to (M,\sG_M), \qquad
\wh\pr_2:(M,\sG_M) \times(M',\sG_{M'})\to (M',\sG_{M'}),
\ee
which one can think of as being trivial superbundles.

\begin{defi} \mar{g36} \label{g36} A (locally trivial) superbundle
over a $G$-supermanifold $\wh M=(M,\sG_M)$ with a typical fiber
$\wh F=(F,\sG_F)$ is a pair $(\wh Y, \wh\pi)$ of a
$G$-supermanifold $\wh Y=(Y,\sG_Y)$ and a $G$-epimorphism
\mar{g37}\beq
\wh\pi: (Y,\sG_Y)\to (M,\sG_M) \label{g37}
\eeq
such that $M$ admits an open cover $\{U_\al\}$ together with a
family of local $G$-isomorphisms
\mar{g38}\beq
\wh\psi_\al: (\pi^{-1}(U_\al),\sG_Y|_{\pi^{-1}(U_\al)})\to
(M,\sG_M|_{U_\al})\times (F,\sG_F). \label{g38}
\eeq
\end{defi}

For any $q\in M$, by $\wh Y_q=\wh\pi^{-1}(q)$ is denoted the
$G$-supermanifold
\mar{g39}\beq
(\pi^{-1}(q), \sG_q=(\sG_Y/{\cal M}_q)|_{\pi^{-1}(q)}),
\label{g39}
\eeq
where ${\cal M}_q$ is a subsheaf of $\sG_Y$ whose sections vanish
on $\pi^{-1}(q)$. This supermanifold is a fiber of $\wh Y$ over
$q\in M$. By a section of the superbundle (\ref{g37}) over an open
set $U\subset M$ is meant a $G$-monomorphism $\wh s: (U,
\sG_M|_U)\to \wh Y$ such that $\wh\pi\circ\wh s$ is the identity
morphism of $(U, \sG_M|_U)$. Given another supermanifold $\wh Y'$
over $\wh M$, a superbundle morphism $\wh\vf:\wh Y\to \wh Y'$ is a
$G$-morphism such that $\wh\pi'\circ\wh\vf=\wh\pi$.

It is readily observed that, by virtue of Definition \ref{g36},
the underlying space $Y$ of a superbundle $\wh Y$ (\ref{g37}) is a
smooth fiber bundle over $M$ with the typical fiber $F$. A section
$\wh s$ of this superbundle defines a section $s$ of the fiber
bundle $Y\to M$, and a superbundle morphism $\wh Y\to\wh Y'$ is
given over a smooth bundle morphism $Y\to Y'$ of their underlying
spaces.

A superbundle over a $G$-supermanifold $\wh M=(M,\sG_M)$ whose
typical fiber is a superspace $(B^{n|m}, \sG_{n|m})$  (seen as the
standard $(n+m,n+m)$-dimensional $G$-supermanifold) is called a
supervector bundle. Transition functions
$\wh\rho_{\al\bt}=\wh\psi_\al\circ\wh\psi_\bt^{-1}$ of a
supervector bundle yield sheaf isomorphisms
\be
\Upsilon_{\al\bt}: \sG_{n|m}|_{U_\al\cap U_\bt}\to
\sG_{n|m}|_{U_\al\cap U_\bt}
\ee
which are described by matrices whose entries are sections of
$\sG_{n|m}|_{U_\al\cap U_\bt}$. Their evaluation
$\dl(\Upsilon_{\al\bt})$ are $GL(n|m;\La)$-valued
$G^\infty$-functions on $U_\al\cap U_\bt$. Sections of a
supervector bundle constitute a sheaf of locally free graded
$\sG_M$-modules. Conversely, let $S$ be a sheaf of locally free
graded $\sG_M$-modules of rank $(r,s)$ on a $G$-manifold $\wh M$,
there exists a supervector bundle over $\wh M$ such that $S$ is
isomorphic to the sheaf of its sections.

For instance, the locally free graded sheaf $\gd \sG_M$ of graded
derivations of $\sG_M$ defines a supervector bundle, called the
tangent superbundle $T\wh M$ of a $G$-supermanifold $\wh M$. If
$(q^1,\ldots,q^{m+n})$ and $(q'^1,\ldots,q'^{m+n})$ are two
coordinate charts on $M$, the Jacobian matrix
\mar{g41}\beq
\Upsilon^i_j=\frac{\dr q'^i}{\dr q^j} \label{g41}
\eeq
(see the prescription (\ref{+83})) provides the transition
function for $T\wh M$. It should be emphasized that the evaluation
$\dl(\Upsilon^i_j)$ of the Jacobian matrix (\ref{g41}) cannot be
written as the Jacobian matrix since odd derivatives of
$G^\infty$-superfunctions are ill-defined.

Turn now to the notion of a principal superbundle with a structure
$G$-Lie supergroup.

Let $\wh e=(e,\La)$ denote a single point with the trivial
$(0,0)$-dimensional $G$-supermanifold structure. For any
$G$-supermanifold, there are natural identifications $\wh
e\times\wh M=\wh M\times\wh e=\wh M$ we refer to in the sequel. A
$G$-supermanifold $\wh K=(K,\sG_K)$ is said to be a $G$-Lie
supergroup if there exist the following $G$-supermanifold
morphisms: a multiplication $\wh m:\wh K\times \wh K\to\wh K$, a
unit $\wh\ve: \wh e\to \wh K$, an inverse $\wh  k:\wh K\to\wh K$,
which satisfy

the associativity $\wh m\circ(\id \times \wh m)=\wh m\circ(\wh
m\times\id):\wh K\times\wh K\times\wh K\to \wh K\times\wh K\to \wh
K$,

the unit property $(\wh m\circ (\wh\ve\times\id))(\wh e\times \wh
K)= (\wh m\circ (\id\times\wh\ve))(\wh K\times \wh e)=\id \wh K$,

the inverse property $\wh m\circ (\wh k\times\id)(\wh K\times\wh
K)=\wh m\circ (\id\times\wh k)(\wh K\times\wh K )=\wh\ve(\wh e)$.

\noindent Given a point $g\in K$, let us denote by $\wh g:\wh
e\to\wh K$ the $G$-supermanifold morphism whose image in $K$ is
$g$. Then one can define the left and right translations of $\wh
K$ as the $G$-supermanifold isomorphisms
\mar{g55}\beq
\wh L_g: \wh K=\wh e\times \wh K\ar^{\wh g\times\id} \wh
K\times\wh K\ar^{\wh m} \wh K,\qquad \wh R_g: \wh K=\wh K\times
\wh e\ar^{\id\times\wh g} \wh K\times\wh K\ar^{\wh m} \wh K.
\label{g55}
\eeq

Apparently, the underlying smooth manifold $K$ of a $G$-Lie
supergroup $\wh K$ is provided with the structure of a real Lie
group, called the underlying Lie group. In particular, the
transformations of $K$ corresponding to the left and right
translations (\ref{g55}) are ordinary left and right
multiplications of $K$ by $g$.

For example, the general linear graded group $GL(n|m;\La)$ is
endowed with the natural structure of an $H^\infty$-supermanifold
of dimension $(n^2+m^2, 2nm)$ so that the matrix multiplication is
an $H^\infty$-morphism. Thus, $GL(n|m;\La)$ is a $H^\infty$-Lie
supergroup. It is trivially extended to the $G$-Lie supergroup
$\wh{GL}(n|m;\La)$, called the general linear supergroup.

A homomorphism $\wh\vf:\wh K\to\wh K'$ of $G$-Lie groups is
defined as a $G$-supermanifold morphism which obeys the conditions
\be
\wh\vf\circ\wh m=\wh m'\circ(\wh\vf\times \wh\vf). \qquad
\wh\vf\circ\wh\ve=\ve', \qquad \wh\vf\circ\wh k=\wh k'\circ\wh\vf.
\ee
In particular, an imbedded monomorphism $\wh H\to\wh K$ of $G$-Lie
supergroups is called a $G$-Lie supersubgroup of $\wh K$.
Accordingly, the monomorphism of underlying Lie groups $H\to K$ is
a Lie subgroup of $K$.

By a right action of a $G$-Lie supergroup $\wh K$ on a
$G$-supermanifold $\wh P$ is meant a $G$-epimorphism $\wh\rho:\wh
P\times \wh K\to \wh P$ such that
\be
\wh\rho\circ (\wh\rho\times\id)=\wh\rho\circ(\id\times\wh m):\wh
P\times \wh K\times \wh K\to \wh P,\qquad \wh\rho\circ (\id\times
\wh \ve)(\wh P\times \wh e)=\id\wh P.
\ee
Apparently, this action implies a right action $\rho$ of the
underlying Lie group $K$ of $\wh K$ on the underlying manifold $P$
of $\wh P$. Similarly, a left action of a $G$-Lie supergroup on a
$G$-supermanifold is defined.

For instance, a $G$-Lie supergroup acts on itself by right and
left translations (\ref{g55}). The general linear supergroup
$\wh{GL}(n|m;\La)$ acts linearly on the standard supermanifold
$\wh B^{n|m}$ on the left by the matrix multiplication which is a
$G$-morphism.

A quotient of the action of a $G$-Lie supergroup $\wh K$ on a
$G$-supermanifold $\wh P$ is a pair $(\wh M, \wh \pi)$ of a
$G$-supermanifold $\wh M$ and a $G$-supermanifold morphism $\wh
\pi:\wh P\to\wh M$ such that:

(i) there is the equality
\mar{+222}\beq
\wh\pi\circ\wh\rho=\wh\pi\circ\wh\pr_1:\wh P\times\wh K\to\wh
M,\label{+222}
\eeq

(ii) for any morphism $\wh\vf:\wh P\to \wh M'$ such that
$\wh\vf\circ\wh\rho=\wh\vf\circ\wh\pr_1$, there is a unique
$G$-morphism $\wh \g:\wh M\to\wh M'$ with $\wh\vf=\wh \g\circ
\wh\pi$. The quotient $(\wh M,\wh\pi)$, denoted by $\wh P/\wh K$,
need not exists. If it exists, its underlying space is $M=P/K$,
and there is a monomorphism of the structure sheaf $\sG_M$ of $\wh
M$ to the direct image $\pi_*\sG_P$. Since the $G$-Lie group $\wh
K$ acts trivially on $\wh M$, the range of this monomorphism is a
subsheaf of $\pi_*\sG_P$, invariant under the action of $\wh K$.
Moreover, there is an isomorphism $\sG_M\cong (\pi_*\sG_P)^{\wh
K}$ of $\sG_M$ to the subsheaf of $\wh K$-invariant sections of
$\sG_P$. The latter is generated by sections of $\sG_P$ on
$\pi^{-1}(U)$, $U\subset M$, which have the same image under the
morphisms
\be
\wh\rho^*:\sG_P|_{\pi^{-1}(U)}\to (\sG_K\wh\ot
\sG_P)|_{(\wh\pi\circ\wh\rho)^{-1}(U)},\qquad
\wh\pr_1^*:\sG_P|_{\pi^{-1}(U)}\to (\sG_K\wh\ot
\sG_P)|_{(\wh\pi\circ\wh\pr_1)^{-1}(U)}.
\ee

\begin{defi} \label{+225} \mar{+225}
A principal superbundle with a structure $G$-Lie supergroup $\wh
K$ is defined as a quotient $\wh\pi: \wh P\to \wh P/\wh K=\wh M$,
which is a locally trivial superbundle with the typical fiber $\wh
G$ such that there exists an open cover $\{U_\al\}$ of $M$
together with $\wh GK$-equivariant trivialization morphisms
\be
\wh\psi_\al: (\pi^{-1}(U_\al),\sG_P|_{\pi^{-1}(U_\al)})\to
(U_\al,\sG_M|_{U_\al})\times \wh K,
\ee
where $\wh K$ acts on itself by left translations.
\end{defi}

Note that, in fact, we need only assumption (i) of the definition
of a quotient and the condition of local triviality of $\wh P$.
Apparently, the underlying smooth bundle $P\to M$ of a principal
superbundle $\wh P\to\wh P/\wh K$ is a principal smooth bundle
with the structure group $K$.

Given a principal superbundle $\wh\pi:\wh P\to\wh M$ with a
structure $G$-Lie supergroup $\wh K$, let $\wh V$ be a
$G$-supermanifold provided with a left action $\varrho: \wh
K\times \wh V\to\wh V$ of $\wh K$. Let us consider the right
action of $\wh K$ on the product $\wh P\times \wh V$ defined by
the morphisms
\mar{g51}\beq
\wh w:\wh P\times \wh V\times \wh K\ar^{\id\times \wh\kappa} \wh
P\times \wh K\times \wh V\ar^{\id\times\wh\Delta\times\id} \wh
P\times \wh K\times \wh K\times \wh
V\ar^{\wh\rho\times\wh\varrho^{-1}} \wh P\times\wh V, \label{g51}
\eeq
where $\wh\kappa:\wh V\times \wh K\to\wh K\times\wh V$ is the
morphism exchanging the factors, $\wh\Delta:\wh K\times\wh K\to\wh
K$ is the diagonal morphism, and
$\wh\varrho^{-1}=\wh\varrho\circ(\wh k\times\id)$. The
corresponding action of the underlying Lie group $K$ on the
underlying smooth manifold $P\times V$ reads
\be
(p\times v)g=(pg\times g^{-1}v), \qquad p\in P,\quad v\in V,\quad
g\in K.
\ee
Then one can show that the quotient $(\wh P\times \wh V)/\wh K$
with respect to the action (\ref{g51}) is a superbundle over $\wh
M$ with the typical fiber $\wh V$. It is called a superbundle
associated to the principal superbundle $\wh P$ because the
underlying smooth manifold $(P\times V)/K$ is a fiber bundle over
$M$ associated to the principal bundle $P$.

\section{The proof of Theorem 2}

Let
\be
\wh\pi:\wh P\to \wh P/\wh K
\ee
be a principal superbundle with a structure $G$-Lie group $\wh K$.
Let $\wh i:\wh H\to \wh K$ be a closed $G$-Lie supersubgroup of
$\wh K$, i.e., $i: H\to K$ is a closed Lie subgroup of the Lie
group $K$. Since $H$ is a closed subgroup of $K$, the latter is an
$H$-principal fiber bundle $K\to K/H$ \cite{ste}. However, $K/H$
need not possesses a $G$-supermanifold structure. Let us assume
that the action
\be
\wh\rho:\wh K\times \wh H\ar^{\id\times \wh i}\wh K\times \wh
K\ar^{\wh m}\wh K
\ee
of $\wh H$ on $\wh K$ by right multiplications defines the
quotient
\mar{g80}\beq
\wh\zeta:\wh K\to \wh K/\wh H \label{g80}
\eeq
which is a principal superbundle with the structure $G$-Lie
supergroup $\wh H$. In this case, the $G$-Lie supergroup $\wh K$
acts on the quotient supermanifold $\wh K/\wh H$ on the left by
the law
\be
\wh\varrho: \wh K\times \wh K/\wh H=\wh K\times \wh\zeta(\wh K)\to
(\wh\zeta\circ\wh m)(\wh K\times \wh K).
\ee

Given this action of $\wh K$ on $\wh K/\wh H$, we have a $\wh
P$-associated superbundle
\mar{g87}\beq
\wh\Si=(\wh P\times \wh K/\wh H)/\wh K \ar^{\wh\pi_\Si} \wh M
\label{g87}
\eeq
with the typical fiber $\wh K/\wh H$. Since
\be
\wh P/\wh H=((\wh P\times \wh K)/\wh K)/\wh H= (\wh P\times \wh
K/\wh H)/\wh K,
\ee
the superbundle $\wh\Si$ (\ref{g87}) is the quotient $(\wh P/\wh
H, \wh\pi_H)$ of $\wh P$ with respect to the right action
\be
\wh\rho\circ (\id\times \wh i):\wh P\times \wh H\ar \wh P\times
\wh K\ar \wh P
\ee
of the $G$-Lie supergroup $\wh H$. Let us show that this quotient
\be
\wh\pi_H:\wh P\to \wh P/\wh H
\ee
is a principal superbundle with the structure supergroup $\wh H$.
Note that, by virtue of the well-known theorem \cite{ste}, the
underlying space $P$ of $\wh P$ is an $H$-principal bundle
\be
\pi_H: P\to P/H.
\ee

Let $\{V_\kappa,\wh\Psi_\kappa\}$ be an atlas of trivializations
\be
\wh\Psi_\kappa : (\zeta^{-1}(V_\kappa),
\sG_K|_{\zeta^{-1}(V_\kappa)})\to
(V_\kappa,\sG_{K/H}|_{V_\kappa})\times \wh H,
\ee
of the $\wh H$-principal bundle $\wh K\to \wh k/\wh H$, and let
$\{U_\al,\wh\psi_\al\}$ be an atlas of trivializations
\be
\wh\psi_\al: (\pi^{-1}(U_\al),\sG_P|_{\pi^{-1}(U_\al)})\to
(U_\al,\sG_M|_{U_\al})\times \wh K
\ee
of the $\wh K$-principal superbundle $\wh P\to \wh M$. Then we
have the $G$-isomorphisms
\mar{g85}\ben
&& \wh\psi_{\al\kappa}=(\id \times \wh\Psi_\kappa)\circ
\wh\psi_\al: (\psi^{-1}_\al(U_\al\times\zeta^{-1}(V_\kappa)),
\sG_P|_{\psi^{-1}_\al(U_\al\times\zeta^{-1}(V_\kappa))})\to  \label{g85}\\
&& \qquad (U_\al,\sG_M|_{U_\al})\times
(V_\kappa,\sG_{K/H}|_{V_\kappa}) \times \wh H=(U_\al\times
V_\kappa, \sG_M|_{U_\al}\wh\ot \sG_{G/H}|_{V_\kappa})\times \wh H.
\nonumber
\een
For any $U_\al$, there exists a well-defined morphism
\be
&& \wh\Psi_\al:(\pi^{-1}(U_\al), \sG_P|_{U_\al}) \to (U_\al\times
G/H, \sG_M|_{U_\al}\wh\ot \sG_{K/H})\times \wh H=\\
&& \qquad (U_\al,\sG_M|_{U_\al})\times \wh K/\wh H\times\wh H
\ee
such that
\be
\wh \Psi_\al|_{\psi^{-1}_\al(U_\al\times\zeta^{-1}(V_\kappa))}
=\wh \psi_{\al\kappa}.
\ee
Let $\{U_\al,\wh\vf_\al\}$ be an atlas of trivializations
\be
\wh\vf_\al:
(\pi_\Si^{-1}(U_\al),\sG_\Si|_{\pi^{-1}_\Si(U_\al)})\to
(U_\al,\sG_M|_{U_\al})\times \wh K/\wh H
\ee
of the $\wh P$-associated superbundle $\wh P/\wh H\to \wh M$. Then
the morphisms
\be
(\wh\vf_\al^{-1}\times \id)\circ \wh\Psi_\al: (\pi^{-1}(U_\al),
\sG_P|_{U_\al})\to (\pi_\Si^{-1}
(U_\al),\sG_\Si|_{\pi^{-1}_\Si(U_\al)})\times\wh H
\ee
make up an atlas
\be
\{\pi^{-1}_\Si(U_\al), (\wh\vf_\al^{-1}\times \id)\circ
\wh\Psi_\al\}
\ee
of trivializations of the  $\wh H$-principal superbundle $\wh P\to
\wh P/\wh H$.

Now, let
\be
\wh i_h : \wh P_h\to \wh P
\ee
be an $\wh H$-principal supersubbundle of the principal
superbundle $\wh P\to\wh M$. Then there exists a global section
$\wh h$ of the superbundle $\wh \Si\to \wh M$ such that the image
of $\wh P_h$ with respect to the morphism $\wh \pi_H\circ\wh i_h$
coincides with the range of the section $\wh h$. Conversely, given
a global section $\wh h$ of the superbundle $\wh \Si\to \wh M$,
the inverse image $\wh\pi_H^{-1}(\wh h(\wh M))$ is an $\wh
H$-principal supersubbundle of $\wh P\to\wh M$.

\section{Supermetrics}

Let us show that, as was mentioned above, the condition of Theorem
\ref{g20} holds if $\wh H$ is the Cartan supersubgroup of a
supermatrix group $\wh K$, i.e., $\wh K$ is a $G$-Lie
supersubgroup of some general linear supergroup
$\wh{GL}(n|m;\La)$.

Recall that a Lie superalgebra $\wh{\got g}$ of an
$(n,m)$-dimensional $G$-Lie supergroup $\wh K$ is defined as a
$\La$-algebra of left-invariant supervector fields on $\wh K$,
i.e., derivations of its structure sheaf $\sG_K$. A supervector
field $u$ is called left-invariant if
\be
(\id\ot u)\circ \wh m^* =\wh m^*\circ u.
\ee
Left-invariant supervector fields on $\wh K$ make up a Lie
$\La$-superalgebra with respect to the graded Lie bracket
\be
[u,u']=u\circ u' -(-1)^{[u][u']}u'\circ u,
\ee
where the symbol $[.]$ stands for the Grassmann parity. Being a
superspace $B^{n|m}$, a Lie superalgebra is provided with a
structure of the standard $G$-supermanifold $\wh B^{n+m,n+m}$. Its
even part $\wh{\got g}_0=\wh B^{n,m}$ is a Lie $\La_0$-algebra.

Let $\wh K$ be a matrix $G$-Lie supergroup. Then there is an
exponential map
\be
\xi(J)=\exp(J)=\op\sum_k \frac{1}{k'}J^k
\ee
of some open neighbourhood of the origin of the Lie algebra
$\wh{\got g}_0$ onto an open neighbourhood $U$ of the unit of $\wh
K$. This map is an $H^\infty$-morphism, which is trivially
extended to a $G$-morphism.

Let $\wh H$ be a Cartan supersubgroup of $\wh K$, i.e., the even
part $\wh{\got h}_0$ of the Lie superalgebra $\wh{\got h}$ of $wh
H$ is a Cartan subalgebra of the Lie algebra $\wh\ccG_0$, i.e.,
\be
\wh\ccG_0=\wh{\got f}_0 +\wh{\got h}_0, \qquad [\wh{\got
f}_0,\wh{\got f}_0]\subset \wh{\got h}_0, \qquad [\wh{\got
f}_0,\wh{\got h}_0]\subset \wh{\got f}_0.
\ee
Then there exists an open neighbourhood, say again $\wh U$, of the
unit of $\wh K$ such that any element $g$ of $\wh U$ is uniquely
brought into the form
\be
g=\exp(F)\exp(I), \qquad F\in \wh{\got f}_0, \qquad I\in \wh{\got
h}_0.
\ee
Then the open set $\wh U_H=\wh m(\wh U\times \wh H)$ is
$G$-isomorphic to the direct product $\xi(\xi^{-1}(U)\cap\wh{\got
f}_0)\times\wh H$. This product provides a trivialization of an
open neighbourhood of the unit of $\wh K$. Acting on this
trivialization by left translations $\wh L_g$, $g\in\wh K$, one
obtains an atlas of a principal superbundle $\wh K\to\wh H$.

For instance, let us consider a superspace $B^{n|2m}$, coordinated
by $(x^a,y^i,\ol y^i)$, and the general linear supergroup
$\wh{GL}(n|2m;\La)$ of its automorphisms. Let $B^{n|2m}$ be
provided with the $\La$-valued bilinear form
\mar{g90}\beq
\om=\op\sum_{i=1}^n (x^i x'^i) + \op\sum_{j=1}^m (y^j\ol y'^j- \ol
y^j y'^j). \label{g90}
\eeq
The supermatrices (\ref{+200}) preserving this bilinear form make
up the orthogonal-symplectic supergroup $\wh{OS}p(n|m;\La)$
\cite{fuks}. It is a Cartan subgroup of $\wh{GL}(n|2m;\La)$. Then
one can think of the quotient
\be
\wh{GL}(n|2m;\La)/\wh{OS}p(n|m;\La)
\ee
as being a supermanifold of $\La$-valued bilinear forms on
$B^{n|2m}$ which are brought into the form (\ref{g90}) by general
linear supertransformations.

Let $\wh M$ be $G$-supermanifold of dimension $(n,2m)$ and $T\wh
M$ its tangent superbundle. Let $L\wh M$ be an associated
principal superbundle. Let us assume that its structure supergroup
$\wh{GL}(n|2m;\La)$ is reduced to the supersubgroup
$\wh{OS}p(n|m;\La)$. Then by virtue of Theorem \ref{g20}, there
exists a global section $h$ of the quotient
\be
L\wh M/\wh{OS}p(n|m;\La)\to \wh M
\ee
which can be regarded as a supermetric on a supermanifold $\wh M$.

Note that, bearing in mind physical applications, one can treat
the bilinear form (\ref{g90}) as {\it sui generis} superextension
of the Euclidean metric on the body $\Bbb R^n=\si(B^{n|m})$ of the
superspace $B^{n|m}$. However, the body of a supermanifold is
ill-defined in general \cite{iva86,caten}.


\begin{thebibliography}{ddd}

\bibitem{ste} N.Steenrod, {\it The Topology of Fibre Bundles} (Princeton Univ.
Press, Princeton, 1972).

\bibitem{sard92} G.Sardanashvily, On the geometry of spontaneous symmetry
breaking, {\it J. Math. Phys.} {\bf 33} (1992) 1546.

\bibitem{higgs} G.Sardanashvily, Geometry of classical Higgs
fields, {\it Int. J. Geom. Methods Mod. Phys.} {\bf 3} (2006)
139-148; {\it arXiv}: hep-th/0510168.

\bibitem{iva} D.Ivanenko and G.Sardanashvily,
The gauge treatment of gravity, {\it Phys. Rep.} {\bf 94} (1983)
1-45.

\bibitem{tmf} G.Sardanashvily, Classical gauge theory of gravity,
{\it Theor. Math. Phys.} {\bf 132} (2002) 1163-1171; {\it arXiv:}
gr-qc/0208054.

\bibitem{west} P.West, {\it Introduction to Supersymmetry and
Supergravity} (Singapore, World Scientific, 1990).

\bibitem{wess} J.Wess and J.Bagger, 1992 {\it Supersymmetry and
Supergravity}, Princeton Series in Physics (Princeton, Princeton
Univ. Press, 1992).

\bibitem{binet} P.Bin\'etry, G.Girardi and R.Grimm, Supergravity
couplings: A geometric formulation, {\it Phys. Rep.} {\bf 343}
(2001) 255.

\bibitem{iva86} D.Ivanenko and G.Sardanashvily, Goldstone type
(non-Poincar\'e) supergravity, {\it Progr. Theor. Phys.} {\bf 26}
(1986) 969.

\bibitem{del} P.Deligne and J.Morgan, Notes on supersymmetry
(following Joseph Bernstein), In: {\it Quantum Field Theory and
Strings: a Course for Mathematicians} (Providence, RI: Amer. Math.
Soc., 1999) p.41.

\bibitem{carm} C.Carmeli, G.Cassinelli, A.Toigo and V.Varadarajan,
Unitary representations of super Lie groups and applications to
the classification and multiplet structure of super particles,
{\it Commun. Math. Phys.} {\bf 263} (2006) 217.

\bibitem{alm} A.Almorox, Supergauge theories in graded manifolds,
In: {\it Differential Geometric Methods in Mathematical Physics},
Lect. Notes in Math. {\bf 1251} (Berlin, Springer, 1987) p.114


\bibitem{boy} C.Boyer and A.Sanchez Valenzuela, Lie supergroup
actions on supermanifolds, {\it Trans. Am. Math. Soc.} {\bf 323}
(1991) 151.

\bibitem{stavr} T.Stavracou, Theory of connections on graded principal
bundles, {\it Rev. Math. Phys.} {\bf 10} (1998) 47.

\bibitem{book00} L.Mangiarotti and G.Sardanashvily, {\it Connections in
Classical and Quantum Field Theory} (World Scientific, Singapore,
2000).

\bibitem{zirn} M.Zirnbauer, Riemannian symmetric superspaces
and their origin in random-matrix theory, {\it J. Math. Phys.}
{\bf 37} (1996) 4986.

\bibitem{aleks} D.Alekseevsky, V.Cort\'es, C.Devchand and
U.Semmelman, Killing spinors are Killing vectors in Riemannian
supergeometry, {\it J. Geom. Phys.} {\bf 26} (1998) 37.

\bibitem{goert} O.Goertsches, Riemannian supergeometry, {\it
arXiv} math.DG/0604143.


\bibitem{bart} C.Bartocci, U.Bruzzo and D.Hern\'andez Ruip\'erez,
{\it The Geometry of Supermanifolds} (Dordrecht, Kluwer Academic
Publ.,1991).


\bibitem{roth} M.Rothstein, The axioms of supermanifolds and a new structure
arising from them, {\it Trans. Amer. Math. Soc.} {\bf 297} (1986)
159.

\bibitem{bart93} C.Bartocci, U.Bruzzo,
D.Hern\'andez Ruip\'erez and V.Pestov, Foundations of
supermanifold theory: the axiomatic approach, {\it Diff. Geom.
Appl.} {\bf 3} (1993) 135.


\bibitem{bruz99} U.Bruzzo and V.Pestov, On the structure of DeWitt
supermanifolds, {\it J. Geom. Phys.} {\bf 30} (1999) 147.

\bibitem{klepp} A.Kleppe and C.Wainwright, Supercost space
geometry, {\it J. Math. Phys.} {\bf 48} (2007) 053511.

\bibitem{rog} A.Rogers, Graded manifolds, supermanifolds and
infinite-dimensional Grassmann algebras, {\it Commun. Math. Phys.}
{\bf 105} (1986) 275.

\bibitem{fuks} D.Fuks, {\it Cohomology of Infinite-Dimensional
Lie Algebras} (Consultants Bureau, N.Y., 1986).

\bibitem{caten} R.Catenacci, C.Reina and P.Teoflatto, On the body of supermanifolds, {\it J.
Math. Phys.} {\bf 26} (1985) 671.

\end{thebibliography}
\end{document}